# Nash bargaining with a nondeterministic threat

Kerry M. Soileau

ABSTRACT

We consider bargaining problems that involve two participants, with a nonempty closed, bounded convex bargaining set of points in the real plane representing all realizable bargains. We also assume that there is no definite threat or disagreement point that will provide the default bargain if the players cannot agree on some point in the bargaining set. However, there is a nondeterministic threat: if the players fail to agree on a bargain, one of them will be chosen at random with equal probability, and that chosen player will select any realizable bargain as the solution, subject to a reasonable restriction.

## 1. INTRODUCTION

We consider bargaining problems that involve two participants, with a nonempty closed, bounded convex <u>bargaining set</u> of points $S \subseteq \mathbb{R}^2$ representing all realizable bargains. That is, to each point corresponds at least one agreement, or bargain, and to each realizable bargain there corresponds a unique point in $S$. The coordinates of each point represent the utility of that bargain to each of the players. We consider the bargaining problem to be equivalent to the selection of some point in $S$ to which both players agree. A solution to the bargaining problem is then regarded as any one of the bargains corresponding to the selected point. We also assume that there is no definite threat or disagreement point $d \in S$ that will provide the default bargain if the players cannot agree on some point in the bargaining set. However, there is a nondeterministic threat: if the players fail to agree on a bargain, one of them will be chosen at random with equal probability, and that chosen player will select any realizable bargain as the solution, with one restriction: there must be no bargain with the same utility to the chosen player, with higher utility to the player not chosen. To put it a different way, if the chosen player can increase the utility to the unchosen player at no personal cost by choosing a different bargain, the chosen player is required to do so. The possibility of satisfying this criterion is assured by the topological compactness of $S$.

## 2. APPROACH

Beginning with the assumption that there exists a solution function that assigns a solution $c(T)$ to every bargaining set $T \subseteq S$, we create a nested diminishing sequence of bargaining sets $\{S_n\}_{n=0}^{\infty}$ for which $c(S_n) = c(S)$, and for which $S_0 = S$. We show $\bigcap_{n=0}^{\infty} S_n$ consists of a single point, and argue that this point is the solution to the original bargaining set $S$, and any of its corresponding bargains is a solution to the original bargaining problem.

## 3. ASSUMPTIONS

Define $t(S) \equiv$ the expected value of the threat point for $S$. Note that if $(x_{\max}, y)$ and $(x, y_{\max})$ are the points in $S$ with maximal abscissa and ordinate, respectively, then

$$t(S) = mid\left((x_{\max}, y), (x, y_{\max})\right) = \left(\tfrac{1}{2}(x_{\max} + x), \tfrac{1}{2}(y + y_{\max})\right).$$

Define $Trim(S) \equiv \{(x, y) \in S ; t(S) \leq (x, y)\}$, where $(a,b) \leq (c,d)$ means $a \leq c$ and $b \leq d$.

We stipulate the following assumptions:

A1) Given the choice between an expected payoff of $x$ and a definite payoff $y$, the player will choose the former if $x > y$, the latter if $x < y$ and will have no preference if $x = y$.

A2) If a bargaining set $T$ contains a bargaining set $S$ and $c(T)$ is in $S$, then

$$c(T) = c(S).$$

A3) If $\{T_n\}_{n=0}^{\infty}$ is a nested sequence of bargaining sets all having the same solution $c(T_0)$, and $\lim_{n \to \infty} Diameter(T_n)$ exists and equals zero, then

$$\{c(T_0)\} = \bigcap_{n=0}^{\infty} T_n.$$

Justification of A1: The players judge a nondeterministic payoff with expected value $x$ and a deterministic payoff of value $x$, as being equally desirable.

Justification of A2: This is precisely assumption 7 in Nash [1950].

Justification of A3: Since $\lim_{n \to \infty} Diameter(T_n) = 0$ and $\bigcap_{n=0}^{\infty} T_n \neq \varnothing$, it follows that $\bigcap_{n=0}^{\infty} T_n$ consists of a single point. Since $c(T_0) \in \bigcap_{n=0}^{\infty} T_n$, $\{c(T_0)\} = \bigcap_{n=0}^{\infty} T_n$.

## 4. A PROPOSITION

Proposition 1: For every bargaining set $T$, $c(T) = c(Trim(T))$.

Proof: First note that because of A1, the first player will never rationally accept any point $(x, y) \in T$ with abscissa less than that of $t(T)$, because in that case the threat scenario has a higher expected payoff to the first player than $(x, y)$ and can be obtained unilaterally by the first player's refusing to bargain. Similarly, because of A1, the second player will never rationally accept any point $(x, y) \in T$ with ordinate less than that of $t(T)$, because in that case the threat scenario has a higher expected payoff to the second player than $(x, y)$ and can be obtained unilaterally by the second player's refusing to bargain. Therefore, the rationality of the players implies $t(T) \leq c(T)$. But this just means that $c(T) \in Trim(T)$. Next, since $T$ contains the set $Trim(T)$ and $c(T) \in Trim(T)$, by A2 we must have $c(T) = c(Trim(T))$.

## 5. A SEQUENCE OF BARGAINING PROBLEMS

We begin with $S_0 = S \subseteq \mathbb{R}^2$ and define $S_n = Trim(S_{n-1})$ for $n \geq 1$. First observe that the sets produced are themselves nonempty closed, bounded convex set of points. They are nonempty because $S_n = Trim(S_{n-1})$ contains the point $t(S_{n-1})$, so by induction since $S_0$ is nonempty, so are all $S_n$. Since

$$Trim(A) = \{(x, y) \in A; t(A) = (t_1, t_2) \leq (x, y)\} = A \cap ([t_1, \infty) \times [t_2, \infty)), \text{ and } [t_1, \infty) \times [t_2, \infty)$$

is closed and convex, if $A$ is closed and convex, so is its intersection with $[t_1, \infty) \times [t_2, \infty)$, thus $Trim(A)$ is closed and convex. Hence by induction since $S_0$ is closed and convex, so are all $S_n$. Finally, since $S_{n+1} \subseteq S_n$ for $n \geq 0$, and $S_0$ is bounded, so are all $S_n$. Thus each $S_n$ is a bargaining set.

Next, by Proposition 1 we have $c(S_n) = c(Trim(S_n))$, so $c(S_n) = c(S_{n+1})$ for every $n \geq 0$. By definition we have $S_{n+1} = Trim(S_n) \subseteq S_n$, so the $\{S_n\}_{n=0}^{\infty}$ are nested. We now show that $\lim_{n \to \infty} Diameter(S_n)$ exists and equals zero. Indeed, for each $S_n$ for $n \geq 1$, let $B_n$ be the smallest closed box such that $S_n \subseteq B_n$. Then we may write $B_n = [a_n, b_n] \times [c_n, d_n]$ for some $a_n, b_n, c_n, d_n \in \mathbb{R}$. If we divide $B_n$ into four congruent disjoint boxes, it is clear that $S_{n+1} = Trim(S_n)$ is a subset of the upper right box $\left[\frac{a_n + b_n}{2}, b_n\right] \times \left[\frac{c_n + d_n}{2}, d_n\right]$.

Indeed, $B_{n+1} \subseteq \left[\frac{a_n + b_n}{2}, b_n\right] \times \left[\frac{c_n + d_n}{2}, d_n\right]$, so $Diameter(B_{n+1}) \leq \frac{1}{2} Diameter(B_n)$. Since $S_1$ is compact, $Diameter(S_1)$ is finite, so is $Diameter(B_1)$, thus $Diameter(B_{n+1}) \leq \frac{1}{2^n} Diameter(B_1)$. Since $Diameter(S_{n+1}) \leq Diameter(B_{n+1})$, we have $Diameter(S_{n+1}) \leq \frac{1}{2^n} Diameter(B_1)$ and thus $\lim_{n \to \infty} Diameter(S_n) = 0$.

Now by A3 we conclude that $\{c(S)\} = \bigcap_{n=0}^{\infty} S_n$. The point $c(S)$, we claim, is the solution to the original bargaining set $S$, and any of its corresponding bargains is a solution to the original bargaining problem.

## 6. AN ALGORITHM FOR FINDING THE SOLUTION

Let $f:[x_{\max}, x] \to [y, y_{\max}]$ be the function such that $\{(t, f(t)); t \in [x_{\max}, x]\}$ is the boundary of $S$ lying between and including $(x_{\max}, y)$ and $(x, y_{\max})$, as defined earlier. Define sequences $\{x_n\}_{n=0}^{\infty}$, $\{y_n\}_{n=0}^{\infty}$, $\{z_n\}_{n=0}^{\infty}$, and $\{w_n\}_{n=0}^{\infty}$ satisfying $x_0 = x$, $y_0 = y_{\max}$, $z_0 = x_{\max}$, and $w_0 = y$, and the recurrences

$x_{n+1} = \frac{1}{2}(x_n + z_n)$, $y_{n+1} = f\left(\frac{1}{2}(x_n + z_n)\right) = f(x_{n+1})$, $z_{n+1} = f^{-1}\left(\frac{1}{2}(y_n + w_n)\right) = f^{-1}(w_{n+1})$

and $w_{n+1} = \frac{1}{2}(y_n + w_n)$ for $n > 0$. Then $c(S) = \lim_{n \to \infty}(x_n, y_n) = \lim_{n \to \infty}(w_n, z_n)$.

## 7. REFERENCES

[1] Nash, John (1950) "The Bargaining Problem" *Econometrica* 18: 155-162.

International Space Station Program Office, Avionics and Software Office, Mail Code OD, NASA Johnson Space Center, Houston, TX 77058
E-mail address: ksoileau@yahoo.com